\documentclass[aps,pre,twocolumn,showpacs,showkeys,a4paper]{revtex4-1}
 \usepackage{graphicx}
 \usepackage{amsmath}
 \usepackage{amssymb}
 \usepackage{amscd}
 \usepackage{color}
 \usepackage{subcaption}

\begin{document} 

 \title{Active dynamics and spatially coherent motion in chromosomes subject
to enzymatic force dipoles}
 \author{Stefanie Put$^{(1)}$, Takahiro Sakaue$^{(2,3)}$ and Carlo Vanderzande$^{(1,4)}$}
 \affiliation{(1) Faculty of Sciences, Hasselt University, 3590 Diepenbeek, Belgium\\
 (2) Department of Physics and Mathematics, Aoyama Gakuin
University, 5-10-1 Fuchinobe, Chuo-ku, Sagamihara, Kanagawa 252-5258,
Japan\\
(3) PRESTO, Japan Science and Technology Agency (JST), 4-1-8
Honcho Kawaguchi, Saitama 332-0012, Japan\\
  (4) Institute for Theoretical Physics, Katholieke Universiteit Leuven, 3001
  Heverlee, Belgium}


 \begin{abstract} 
Inspired by recent experiments on chromosomal dynamics, we introduce an exactly solvable model for the interaction between a flexible polymer and a set of  motor-like enzymes. The enzymes can bind and unbind to specific sites of the polymer and when bound produce a dipolar force on two neighboring monomers. We study the resulting non-equilibrium dynamics of the polymer and find that the motion of the monomers has several properties that were observed experimentally for chromosomal loci:   a subdiffusive mean squared displacement and the appearance of regions of correlated motion. We also determine the velocity autocorrelation of the monomers and find that the underlying stochastic process is not fractional Brownian motion. Finally, we show that the active forces swell the polymer by an amount that becomes constant for large polymers.
\end{abstract}

\maketitle 
\section{Introduction}
Within a cell there exist several motorlike enzymes that exert forces on their substrate thereby putting it out of equilibrium. We can think about myosin that contracts actin filaments in the cytoskeleton \cite{Howard01},  helicases and topoisomerases that unwind or change the topology of DNA, chromatin remodeling complexes that allow relocation of nucleosomes \cite{Narlikar13} or chaperones that stabilise the native form of proteins \cite{Goloubinoff18}. In each of these cases, the enzyme uses the free energy from ATP hydrolysis to produce forces that act on the underlying biopolymer, thereby modifying its structure and dynamics. 
The enzyme-substrate complex can be seen as a form of active matter as it converts free energy into motion \cite{Marchetti13}. 

As an example of the resulting non-equilibrium behavior, we mention several recent experiments on the motion of chromosomal loci in bacteria and eukaryotes \cite{Weber12,Stadler17,Zidovska13,Saintillan18}. Evidence has been found for ATP-dependent enhanced (sub)diffusion and coherent motion. These occur on time scales that are of the order of seconds, which are the typical times on which active proteins work. 

In this paper we investigate in a simple model how the action of active enzymes modifies the dynamics of a polymer. 
We model the latter as a flexible polymer (Rouse model). Such a description is appropriate for DNA on length scales above its persistence length. Besides the polymer we have a set of enzymes that can bind/unbind to the substrate, either at all monomers, or at a specified subset of them. When bound, the enzyme exerts a dipolar force on the polymer. The appearance of force dipoles is quite common in active systems. Several active enzymes, such as the chromatin remodeling complexes, produce such forces on DNA.

The work presented here is a continuation of earlier studies in which various authors investigated the effect of monopolar active forces on the dynamics of a flexible or semiflexible polymer \cite{Bruinsma14,Gosh14,Kaiser15,Vandebroek15,Sakaue16,Samanta16,Osmanovic17,Eisenstecken17,Winkler17,Osmanovic18}.

Within our model it is possible to calculate standard polymer properties such as the average squared end-to-end distance and the mean squared displacement of individual monomers. We also pay particular attention to correlation functions such as those of the position and the velocity of the monomers. 
In general we find a good qualitative agreement with dynamical phenomena observed in chromatin. For example we show that the monomers perform a subdiffusion, as found in experiments. This is in contrast with earlier work on polymers subject to active  monopolar forces where the monomers showed a superdiffusive motion. We also observe that the active dipoles lead to the appearance of regions with correlated motion. Finally, we find that the squared end-to-end distance of the chain increases with an amount that is independent of the length of the polymer. 

This paper is organised as follows. In section II we define our model. In section III we study the dynamics of the Rouse modes. For active forces that are heterogeneous along the chain, the modes become coupled, a signature of true non-equilibrium behavior \cite{Battle16,Gladrow16}. In section IV we determine the motion of a single monomer and show that it is subdiffusive. We also study the velocity autocorrelation function and find that on the time scales of the active forces it deviates significantly from the thermal one. In section V we go deeper into the difference between monopolar and dipolar active forces. Using an argument based on the propagation of stress in a Rouse chain, we show that the former give rise to superdiffuse behavior, whereas the latter lead to subdiffusive motion.  In section VI we present our results for the correlation in the motion between different monomers while in section VII we study the mean squared end-to-end distance of the chain. Finally in section VIII,  we present our conclusions and their possible biological relevance. 

\section{Model}
Our model describes the interaction between two (bio)polymers: the first one is a long flexible polymer (e.g. DNA on length scales above its persistence length) with $N$ (assumed to be even) monomers at positions $\vec{R}_n(t)$ ($n \in {\cal N}=\{0, \dots, N-1\}$). Secondly, we have a reservoir with $M$ active enzymes (here we take $M>N/2$) which can bind (unbind) to the polymer with rate $\lambda_o$ ($\lambda_f$). An enzyme binds on two consecutive monomers (say $n$ and $n+1$) and, when bound, exerts a dipolar force on the polymer, i.e. when monomer $n$ feels a force $\vec{F}_n$, monomer $n+1$ experiences a force $-\vec{F}_n$. These forces are assumed to be produced by the hydrolysis of ATP. To be concrete, one can think of active enzymes such as helicases, topoisomerases or chromatin remodeling complexes to produce such a dipolar force on chromatin. 

\begin{figure}[h]
\centering
\includegraphics[width=8cm]{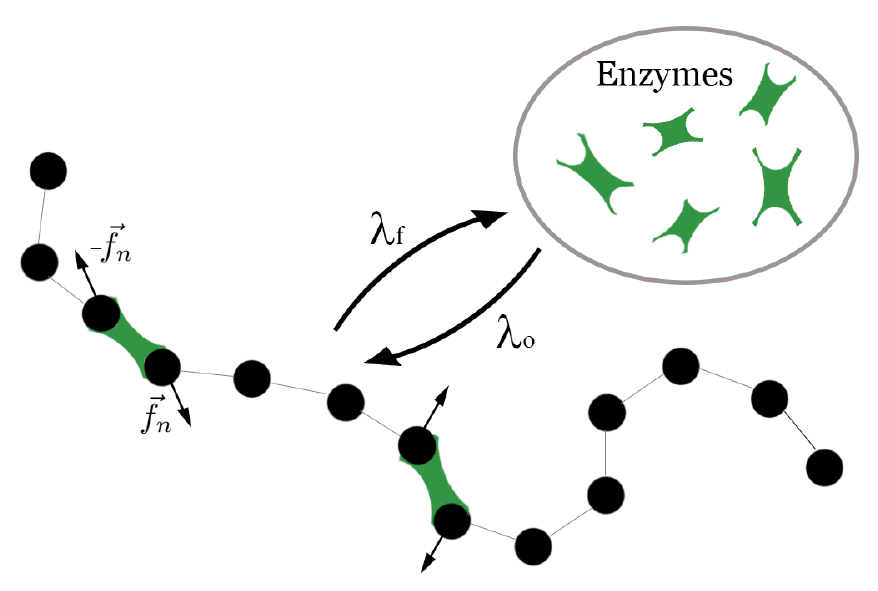}
\caption{Schematic representation of the model. Enzymes (green) can bind (unbind) to a long polymer chain (black) and when bound produce a dipolar force.}
\label{fig1}
\end{figure}

A schematic representation of our model is given in Fig. \ref{fig1}. We distinguish between the cases where the enzymes can bind to all monomers, and that where they can only bind on a sparse subset (typically ten percent) thus producing a heterogeneity along the chain. We will denote the set  of monomers on which the dipoles can bind by ${\cal \nu} \subseteq {\cal N}$. 

In order to get a simple, solvable model we will neglect effects of self-avoidance and assume that the monomers are bound through an harmonic potential with spring constant $k$. Their dynamics then becomes that of the Rouse model \cite{Rouse55,Doi86}. In an overdamped regime, the equation of motion of the $n$-th monomer is given by the Langevin equation
\begin{multline}\label{EQ:GLE}
\gamma \dfrac{d\vec{R}_n(t)}{dt} = -k\left(2\vec{R}_n(t)-\vec{R}_{n-1}(t)-\vec{R}_{n+1}(t)\right) \\ + \vec{\xi}_{T,n}(t) + \vec{f}_n(t) H(t),
\end{multline}
where $\gamma$ is the friction constant. $H(t)$ is the Heaviside function, i.e. we assume that at $t=0$ the model is in thermal equilibrium after which we turn on the active forces (by e.g. adding ATP to the system).   The thermal noise $\vec{\xi}_{T,n}(t)$ is assumed to be Gaussian with zero average and correlation given by the fluctuation-dissipation theorem
\begin{equation}
\langle \vec{\xi}_{T,n}(t)\vec{\xi}_{T,m}(t') \rangle = 6k_BT \gamma \delta_{n,m} \delta(t-t')
\label{fdt}
\end{equation}

Given the substrate-enzyme interaction introduced above, the force exerted by the enzyme on the $n$-th monomer can be written as
\begin{equation}
\vec{f}_n(t) =\vec{F}_n \chi_n(t) I_{n \in {\cal \nu}}
\end{equation}
where $\chi_n(t)$ is a dichotomous Markov process with rates $\lambda_o$ en $\lambda_f$ and $I_{n \in {\cal \nu}}$ is an indicator function which is $1$ ($0$) when the monomer $n$ is (not) in the set ${\cal \nu}$. 
In biological systems, one can expect that the dipolar forces have a specified direction, for example they can be oriented along the vector separating the neighboring monomers. For such a force it is however not possible to obtain an exact solution to our model. Moreover, it could be argued that if one describes the chromosome at some coarse grained scale $b$, the activity of several enzymes in the volume $b^3$ can appear as random in direction. For these reasons, we assume that the force $\vec{F}_n$ it is randomly oriented with independent components that are taken from a uniform distribution on $[-f,f]$. One then has $\langle \vec{f}_n(t) \rangle=0$ and
\begin{equation}
\langle \vec{f}_n(t) \vec{f}_m(t') \rangle = \dfrac{f^2\lambda_\textnormal{o}}{\lambda_\textnormal{o} + \lambda_\textnormal{f}}e^{-\lambda_\textnormal{f} |t'-t|} (\delta_{n,m}-\delta_{n+1,m}) I_{n \in {\cal \nu}}.
\label{3}
\end{equation}
(This relation holds for $n$ even. For $n$ odd, $\delta_{n+1,m}$ has to be replaced by $\delta_{n-1,m}$.)
In this way the (active) force produced by the enzymes has an exponential correlation in time as often assumed in previous models of polymers subject to active forces. It is also important to remark that $\vec{f}_n(t)$ is non-Gaussian \cite{Chaki18} and therefore we can expect that also the motion of the monomers will become non-Gaussian. However, in the present study we only study second moments and two-point correlations, quantities where non-Gaussianity cannot be observed. 

The Langevin equations (\ref{EQ:GLE}) have to be solved with the appropriate boundary conditions which for an open chain are given in terms of the position of two ghost monomers with $n=-1$ and $n=N$: $\vec{R}_{-1}(t)=\vec{R}_0(t),\ \vec{R}_N(t)=\vec{R}_{N-1}(t)$. 

Solving (\ref{EQ:GLE}) allows us to study the response of the polymer to the addition of, for example, ATP. We expect that asymptotically in time, the solution will go to a new non-equilibrium steady state (NESS). In the case of a homogeneous model (enzymes can bind everywhere) we expect that this NESS will be similar to the equilibrium one since on time scales much larger than $1/\lambda_f$ the correlations of the active forces are delta-correlated like the thermal ones. The NESS will therefore be similar to the equilibrium state, albeit at a higher effective temperature. For a heterogeneous system (enzymes can bind at a subset of monomers) the same reasoning shows that in the NESS the system can be seen as one with a non-homogeneous temperature, leading to true non-equilibrium behavior \cite{Falasco15}.

\section{Dynamics of Rouse modes}
The set of Langevin equations (\ref{EQ:GLE}) can be solved by a standard transformation to normal coordinates $\vec{X}_p(t)$ ($p \in \{0,\ldots,N-1\}$), in the polymer context often referred to as Rouse modes \cite{Doi86}: 
\begin{equation}
\vec{X}_p(t) = \dfrac{1}{N} \sum_{n=0}^{N-1}C^p_n\vec{R}_n(t),
\end{equation}
with coefficients
\begin{equation}
C^p_n = \textnormal{cos}\left(\dfrac{\pi p}{N}(n+\dfrac{1}{2})\right).
\end{equation}
For $p=0$, the coordinate is equivalent to the location of the centre of mass ($\vec{R}_{\textnormal{CM}}$).
One can go back to the original coordinates with the reverse transformation
\begin{align}
\vec{R}_n(t) = \vec{X}_0(t) + 2\sum_{p=1}^{N-1}C^p_n\vec{X}_p(t).
\label{monpos}
\end{align}
A straightforward calculation shows that the Rouse modes obey the equations
\begin{align}
\gamma \frac{d\vec{X}_p(t)}{dt} = -\dfrac{k_p}{2N}\vec{X}_p(t)+\vec{\xi}_{T,p}(t)+\vec{f}_p(t),
\label{LR}
\end{align}
These are Langevin equations for a particle moving in a harmonic potential with spring constant 
\begin{equation}\label{eq:kp}
k_p = 8Nk\ \textnormal{sin}^2\left(\dfrac{\pi p}{2N}\right) \approx \dfrac{
2k\pi^2 p^2}{N}.
\end{equation}
where the approximation is valid for large $N$. The particle is subject to a normal thermal
noise $\vec{\xi}_{T,p}(t)$ 
\begin{equation}
\vec{\xi}_{T,p}(t) = \dfrac{1}{N}\sum_{n=0}^{N-1}C^p_n\vec{\xi}_{T,n}(t).
\end{equation}
and a normal active force 
\begin{equation}
\vec{f}_{p}(t) = \dfrac{1}{N}\sum_{n\in \nu}C^p_n\vec{f}_{n}(t).
\end{equation}

The normal thermal force $\vec{\xi}_{T,p}(t)$ is again a Gaussian random variable with zero average and a correlation which can be obtained from (\ref{fdt}). It equals 
\begin{equation}
\langle \vec{\xi}_{T,p}(t)\vec{\xi}_{T,q}(t') \rangle = \dfrac{3k_BT}{2N} \gamma (1+\delta_{p,0})\delta_{p,q} \delta(t-t'),
\label{thermalcor}
\end{equation}
where we have used the orthogonality relation
\begin{equation}\label{eq:orthoCoef}
\sum_{n=0}^{N-1}C^p_nC^q_n = \dfrac{N}{2}\left(1+\delta_{p,0}\right)\delta_{p,q}.
\end{equation}

While the average of the normal active noises $\vec{f}_{p}(t)$ is also zero, their correlation is more complicated. Using (\ref{3}) we find
\begin{multline}
\langle \vec{f}_p(t) \vec{f}_q(t') \rangle = \dfrac{1}{N^2}\sum_{n \in \nu}\sum_{m \in \nu}C^p_n C^q_m \langle \vec{f}_n(t) \vec{f}_m(t') \rangle \\ = \dfrac{f^2}{N^2}\dfrac{\lambda_\textnormal{o}}{\lambda_\textnormal{o} + \lambda_\textnormal{f}}e^{-\lambda_\textnormal{f} |t'-t|} G_{p,q}
\label{activcor}
\end{multline}
with 
\begin{equation}
G_{p,q}=\sum_{n\ \textnormal{even} \atop n \in {\cal \nu}} \left( C_n^p - C_{n+1}^p \right) \left( C_n^q - C_{n+1}^q \right)
\end{equation}
In general $G_{p,q}$ is not proportional to $\delta_{p,q}$, not even when the enzymes act on all the monomers. This leads to a coupling of the different active noise contributions $\vec{f}_p(t)$, the consequences of which will be discussed below.


The Langevin equation (\ref{LR}) for the normal coordinates can easily be solved with the result\begin{eqnarray}
\vec{X}_p(t)&=&\vec{X}_p(0)e^{-\frac{t}{\tau_p}} + \dfrac{1}{\gamma}\int_0^t \vec{\xi}_{T,p}(t-\tau)e^{-\tau/\tau_p} d\tau \nonumber \\ &+& \dfrac{1}{\gamma}\int_0^t \vec{f}_{T,p}(t-\tau)e^{-\tau/\tau_p} d\tau.
\label{normalmode}
\end{eqnarray}
Here, we introduced the characteristic time for the p$^{th}$ mode
\begin{equation}\label{eq:taup}
\tau_p = \dfrac{2\gamma N}{k_p} \ \ \ \ \  (p \neq 0).
\end{equation}
The largest of these relaxation times, $\tau_1$, is referred to as the Rouse time. Physically, it corresponds to the time that the polymer takes to diffuse over a distance of the order of its radius of gyration. We also have $\tau_0=0$, in which case the exponentials in (\ref{normalmode}) are equal to $1$.

We can now determine the statistical properties of the Rouse modes which will be used in the next section to determine those of the monomers. However, as recent work shows, the motion of these modes in itself has interesting properties for systems that are out of equilibrium \cite{Battle16,Gladrow16}. 

Clearly, the average of each Rouse mode relaxes to zero, while the correlation between two modes equals 
\begin{multline}\label{eq:corRouseModes}
\langle\vec{X}_p(t)\vec{X}_q(t')\rangle = \langle\vec{X}_p(0)\vec{X}_q(0)\rangle e^{-t/\tau_p} e^{-t'/\tau_q} \\ + \dfrac{1}{\gamma^2}\int_0^td\tau \int_0^{t'}d\tau'\Big(\langle\vec{\xi}_{T,p}(t-\tau)\vec{\xi}_{T,q}(t'-\tau')\rangle \\ + \langle\vec{f}_{p}(t-\tau)\vec{f}_{q}(t'-\tau')\rangle\Big) e^{-\tau/\tau_p} e^{-\tau'/\tau_q},
\end{multline}
where we used the fact that the initial value of the normal coordinates, the normal thermal noise and the normal active noise are uncorrelated. 

Since the system is in equilibrium at $t=0$ the equipartition theorem can be applied to the normal modes. Inserting (\ref{thermalcor}) and (\ref{activcor}) into (\ref{eq:corRouseModes}) we obtain\begin{multline}\label{eq:XpXq}
\langle\vec{X}_p(t)\vec{X}_q(t')\rangle = \dfrac{3k_B T}{k_p} e^{- |t-t'|/\tau_p}\ \delta_{p,q} \\ + \dfrac{f^2\lambda_\textnormal{o}}{N^2\gamma^2(\lambda_\textnormal{o}+\lambda_\textnormal{f})}
 G_{p,q} H_{p,q}(t,t'),
\end{multline}
where
\begin{multline}
H_{p,q}(t,t') = \int_0^t\int_0^{t'} d\tau d\tau' e^{-\lambda_\textnormal{f}|t-t'-\tau+\tau'|} e^{-\tau/\tau_p} e^{-\tau'/\tau_q}.
\label{integrals}
\end{multline}
These integrals are easy to calculate so we don't give their lengthy expressions here.

\begin{figure}
\centering
\includegraphics[width=8cm]{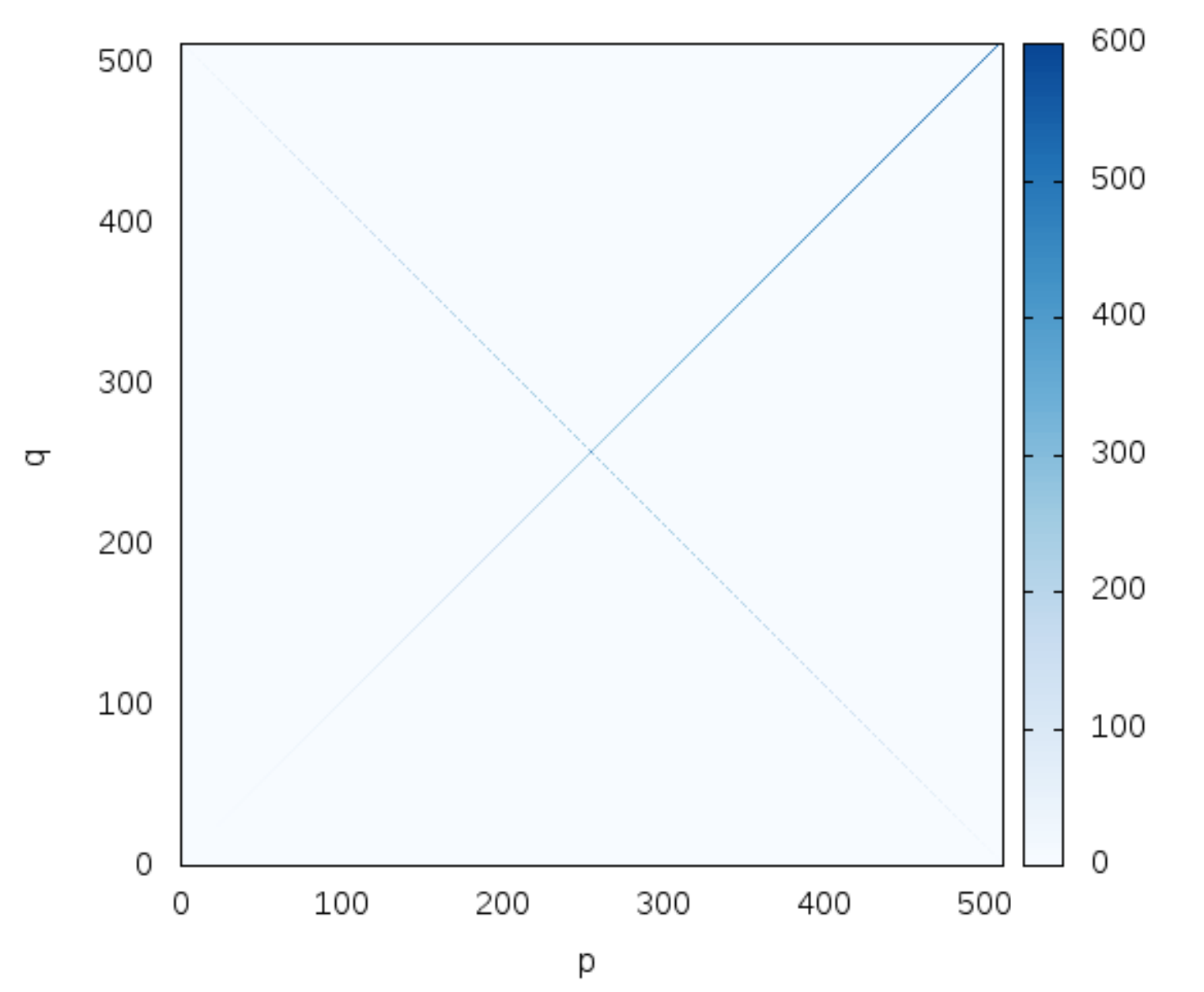}
\includegraphics[width=8cm]{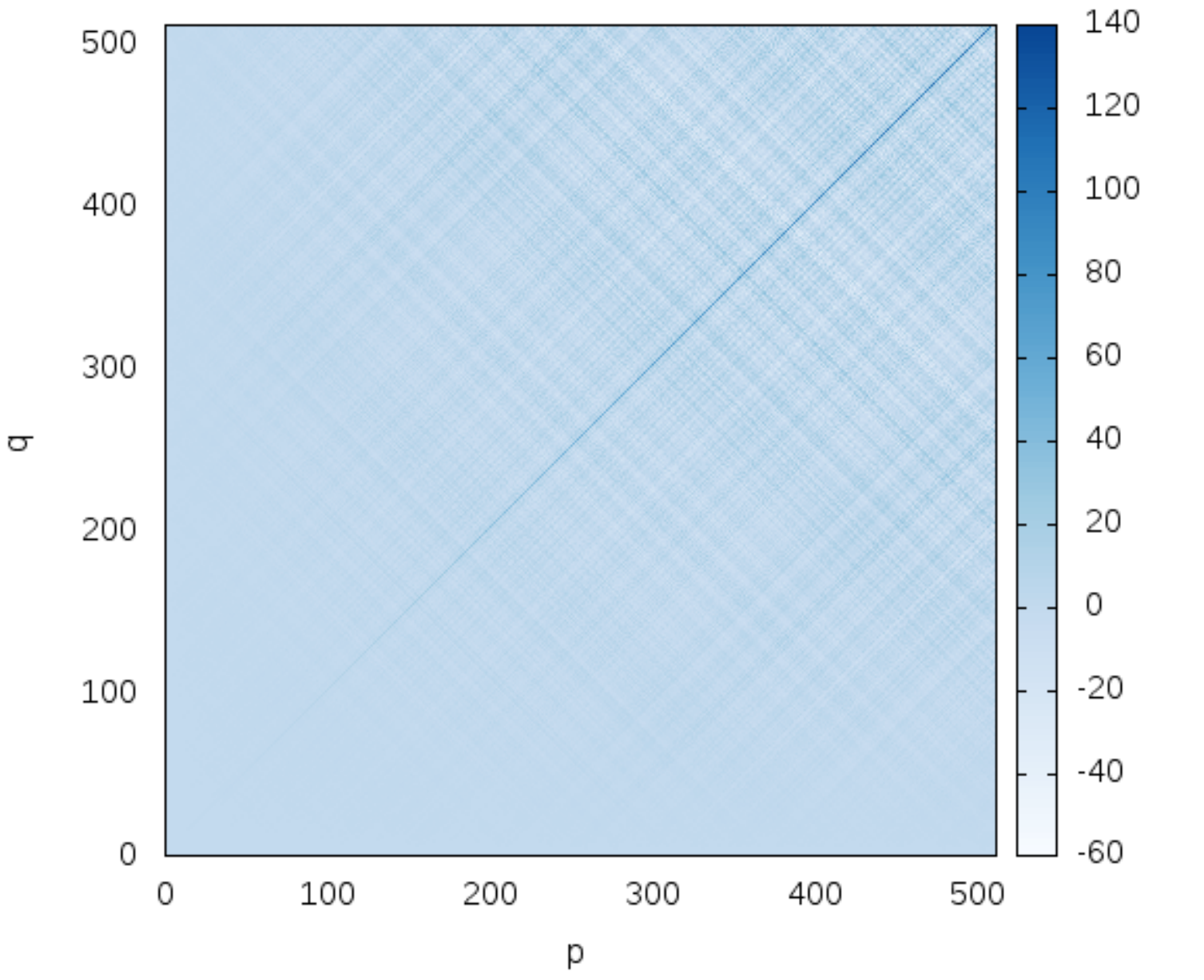}
\caption{\label{fig:FGpq} Matrix $G_{p,q}$ for active forces on all monomers (top) and 
for forces on 51 randomly chosen pairs of monomers (bottom). The number of monomers is $512$. }
\label{fig2}
\end{figure}

We notice an important difference between the effect of the thermal forces and the active ones. The first ones are uncorrelated and act on all monomers and as a consequence do not lead to a correlation between the Rouse modes. The dipolar active forces are themselves correlated and (for the case of heterogeneous active forces) do not act on all the monomers. Both ingredients are encoded in the matrix $G_{p,q}$ which is not diagonal. Indeed, for the case that the active force can act on all the monomers (${\cal \nu}={\cal N}$) it is possible to show (see supplemental material) that
\begin{multline}\label{eq:Gpq}
G_{p,q} = \dfrac{N}{2}\left(1-\textnormal{cos}\left(\dfrac{\pi p}{N}\right)\right)\delta_{p,q} \\ + \dfrac{N}{2}\textnormal{cos}\left(\dfrac{\pi (q-p)}{2N}\right)\delta_{p+q,N}.
\end{multline}
Each mode gets coupled with one other mode. This effect can easily be understood since the dipolar interactions decrease the periodicity of the chain by a factor two.
A representation of the matrix $G_{p,q}$ for that situation is shown in Fig. \ref{fig2} (top) for a polymer with $N=512$. 
In general it is not possible to obtain analytical expressions for $G_{p,q}$. We show in Fig. \ref{fig2} (bottom) an example of $G_{p,q}$ for the case where there are 51 randomly chosen pairs of sites on which the enzymes can bind (again in a polymer with 512 monomers). As can be seen, there appears a coupling between a large number of modes. This coupling leads to a current in the phase space of normal modes, a clear signature of the breaking of detailed balance \cite{Battle16,Gladrow16}. We will not go into these issues but  focus here on the motion of the monomers and the similarity with the dynamical properties of chromatin. 

\section{Subdiffusive monomer motion}
We now turn our attention to the motion of the monomers. Specifically we determine their mean squared displacement and the related velocity autocorrelation function. 

As already mentioned, we start at $t=0$ from a thermal equilibrium state and then calculate the response as the active forces are turned on. After a long time, the polymer will reach a new non-equilibrium steady state (NESS). Experimentally, this situation can be studied by starting from a situation where the production of ATP is inhibited. After some time, the drugs that deplete ATP are removed so that its production starts again. Experiments like these are now carried out routinely. 

When a force is acting on a given monomer, its effects will spread through the whole polymer through stress propagation. It has been shown that in a Rouse chain this is a diffusive process \cite{Sakaue12}. Therefore, it takes a time of the order of $N^2$ for the effect of the force to spread through the whole chain. For this reason we expect that the NESS will be reached in a time that is of the order of the Rouse time, an expectation that is in agreement with the results discussed below. This argument holds when the Rouse time is much larger than the correlation time of the active forces. In biological systems this relation should always hold.

\subsection{Mean squared displacement}

We introduce the quantity $\Delta \vec{R}_n(t)= \vec{R}_n(t)-\vec{R}_n(0)$ that corresponds to the motion of the $n$-th monomer in time $t$. The mean squared displacement of the $n$th monomer, $d_n^2(t)$ is given by $ \langle \Delta \vec{R}_n(t)\cdot \Delta \vec{R}_n (t) \rangle$. 

Using (\ref{monpos}) and (\ref{eq:XpXq}), one can easily find that 
\begin{eqnarray}
d_n^2(t)&=&\frac{6 k_B T}{\gamma N} t  + 24k_B T \sum_{p=1}^{N-1} (C_n^p)^2  \left(1- \exp(-t/\tau_p)\right)/k_p \nonumber \\
&+& \frac{4 \lambda_0 f^2}{\gamma^2 N^2 (\lambda_0+\lambda_f)} \sum_{p=1}^{N-1}\sum_{q=1}^{N-1} C_n^p C_n^q G_{p,q} H_{p,q}(t,t) \nonumber \\
\label{corr}
\end{eqnarray}
The first term corresponds to the thermal diffusion of the center of mass with a diffusion constant inversely proportional to $N$, a contribution well known from equilibrium polymer physics. The second and third term correspond to the fluctuations in monomer position around that diffusion caused respectively by thermal and active fluctuations. Notice that, as should be the case, the dipolar active forces do not contribute to the centre of mass motion. 

The time dependence of the total thermal contribution is well known  \cite{Doi86}. After an initial regime in which a monomer performs a free diffusion, its motion becomes subdiffusive, $d_n^2(t) \sim t^{1/2}$. Finally, on time scales above the Rouse time, the second contribution in (\ref{corr}) becomes constant and the monomer follows the motion of the whole polymer as given by the center of mass term. 

How is this behavior modified by the active forces? The latter introduce a new time scale $\tau_f=1/\lambda_f$ which we expect to be of the order of seconds. For a long chain we then have the following relation between the various time scales,
$\tau_N \ll \tau_f \ll \tau_1$. 

For very small times where $t$ is smaller than $\tau_f$ and all the relaxation times of the Rouse modes, the integrandums in (\ref{integrals}) are constant and therefore one finds $d_n^2(t) \sim t^{2}$. This is the ballistic behavior of a particle subject to a constant force that is not yet aware that it is part of a polymer. On the other hand, for very large times, $t \gg \tau_1$, the integrals (\ref{integrals}) saturate and therefore the active contribution becomes a constant. The monomer then performs ordinary diffusion.

\begin{figure}[h]
\centering
\includegraphics[width=8cm]{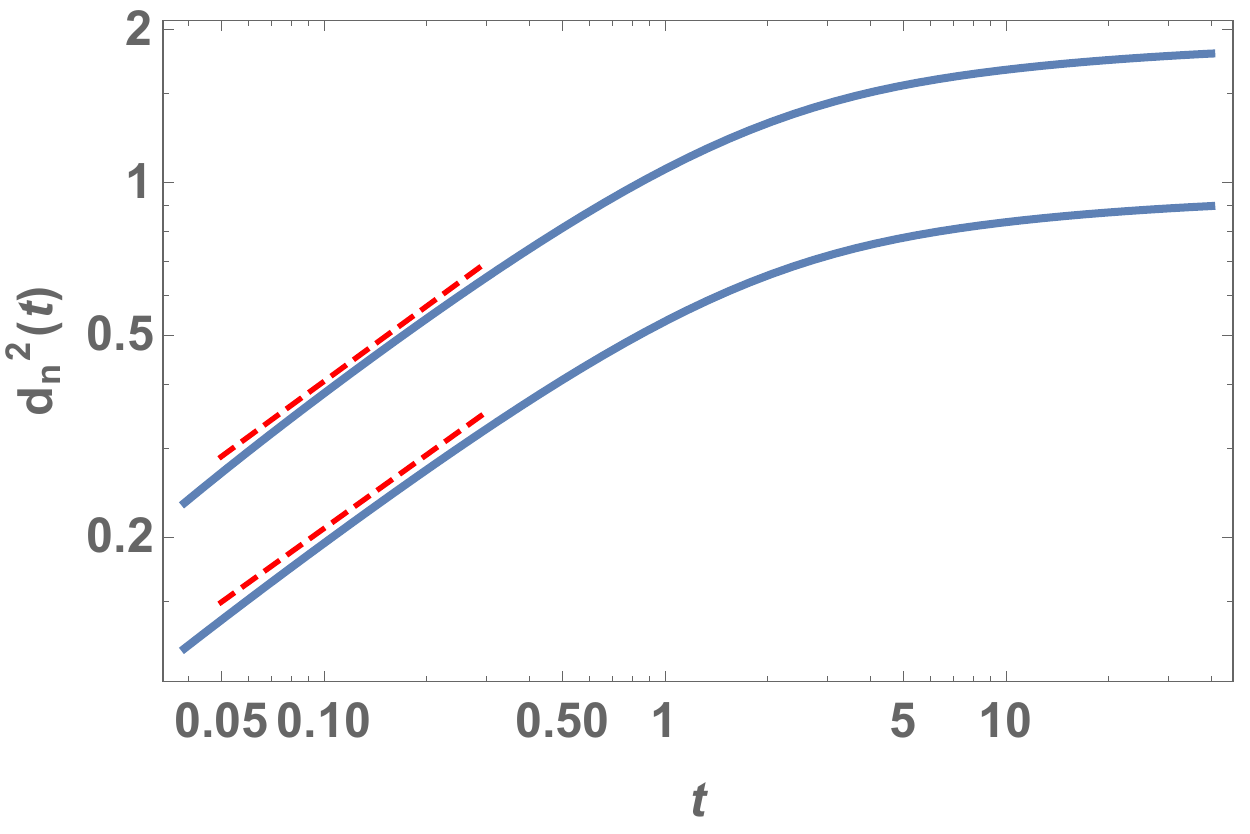}
\caption{Log-log plot of the mean squared displacement $d_n^2(t)$ (only the active contribution is shown) for the monomer with $n=255$ (lower curve) and with $n=511$ (upper curve) in a chain with $N=512$. Enzymes can bind to all pairs of monomers. The dashed lines represent a best fit through the data for $0.05<t<0.30$ and have a slope of $0.48$ respectively $0.49$. The mean squared displacement is measured in $\mu$m$^2$, the time in seconds.}
\label{fig4}
\end{figure}

The intermediate time regime in which $\tau_N \ll t \ll \tau_f$ is the most interesting one. For the case where enzymes can bind on all monomers and an exact expression for $G_{p,q}$ is known, it can be shown from (\ref{corr}) (see supplemental material) that in this regime the monomers subdiffuse and $d_n^2(t) \sim t^{1/2}$, i.e. have the same behavior as caused by thermal forces. A similar conclusion can be drawn from stress propagation arguments (see next section). In Fig. \ref{fig4} \cite{figures}, we have plotted the exact result (\ref{corr}) for a monomer near the middle and at the end of a polymer with $N=512$ and $\tau_f=1$ second. The predicted subdiffusive behavior is indeed recovered.

\begin{figure}
\centering
\includegraphics[width=8cm]{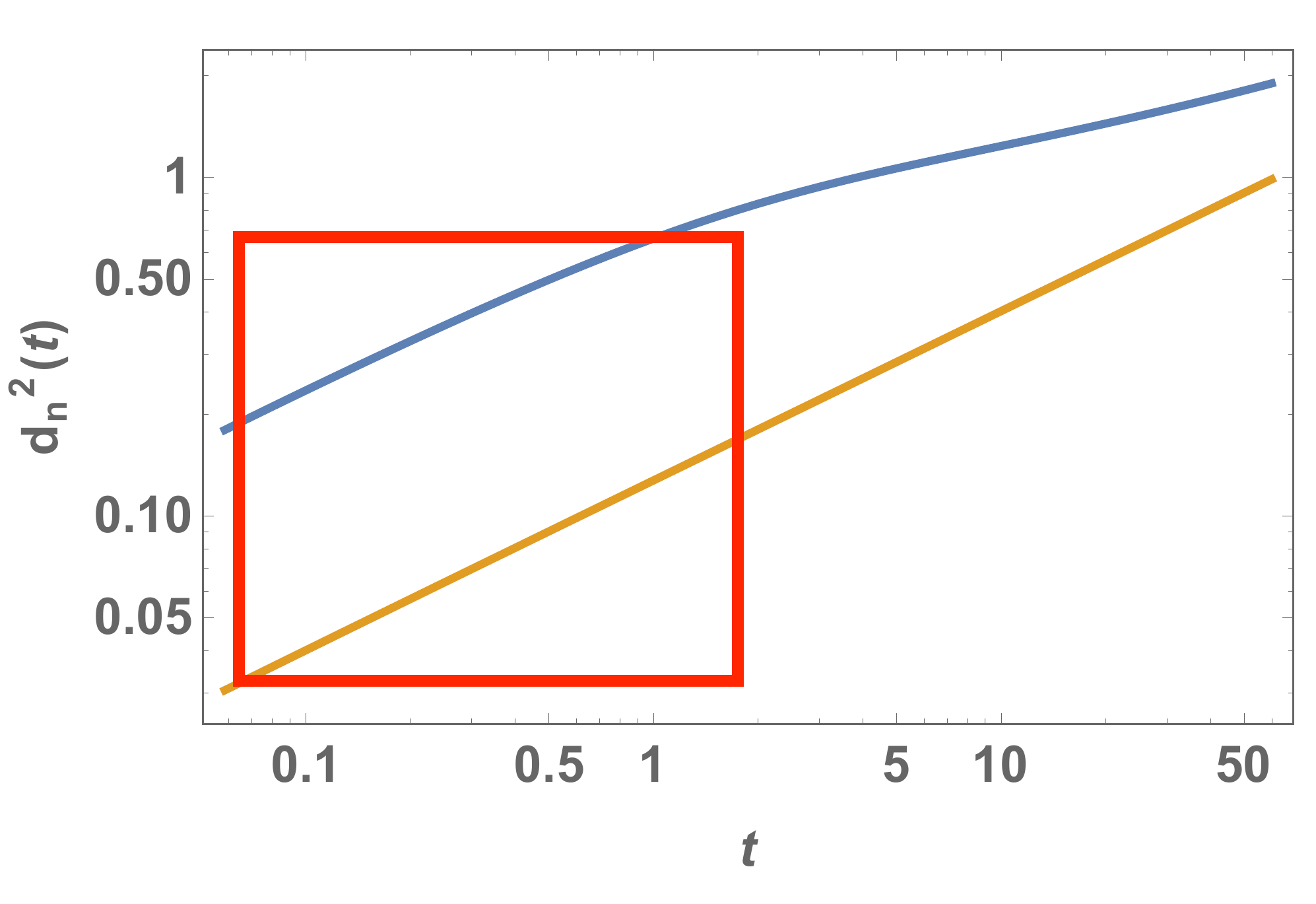}
\caption{Total (thermal + active) mean squared displacement of a monomer (upper curve) and thermal contribution (lower curve) as a function of time. Within the red square the amplitude of both curves is different but their slope is the same.}
\label{fig4b}
\end{figure}

The active contribution has to be added to the thermal one to obtain the total mean squared displacement. For times smaller than $\tau_f$ this leads to a subdiffusion with the same exponent as in the purely thermal case, but with an enhanced amplitude (see Fig. \ref{fig4b}). This is similar to what was observed for the motion of loci in the chromosomes of {\it E. Coli} where it was found that inhibiting the production of ATP decreased the amplitude but didn't change the exponent of the observed subdiffusion \cite{Weber12}. A similar decrease of the diffusion constant was recently observed in the motion of telomeres \cite{Stadler17}. For times larger than $\tau_f$ the effect of the active forces becomes constant and the motion crosses over into the thermal one.

\begin{figure}
\centering
\includegraphics[width=4cm]{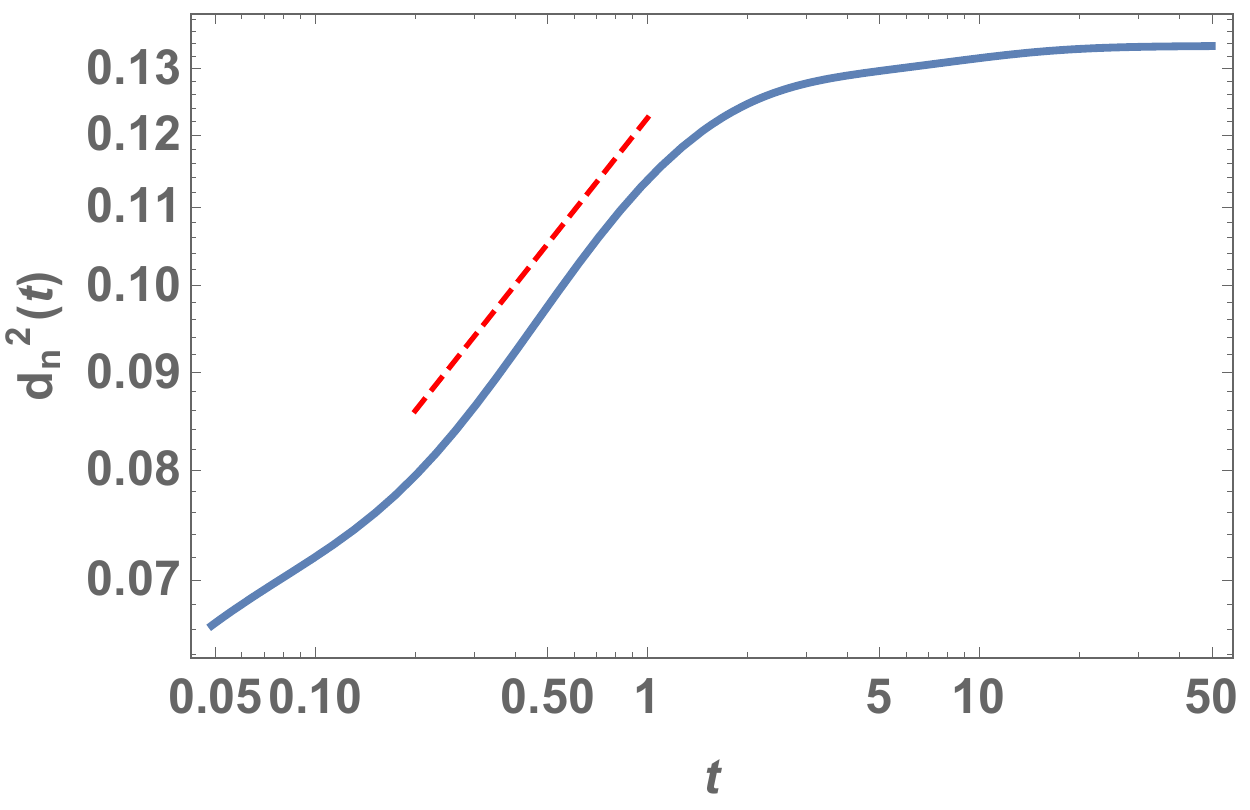}
\includegraphics[width=4cm]{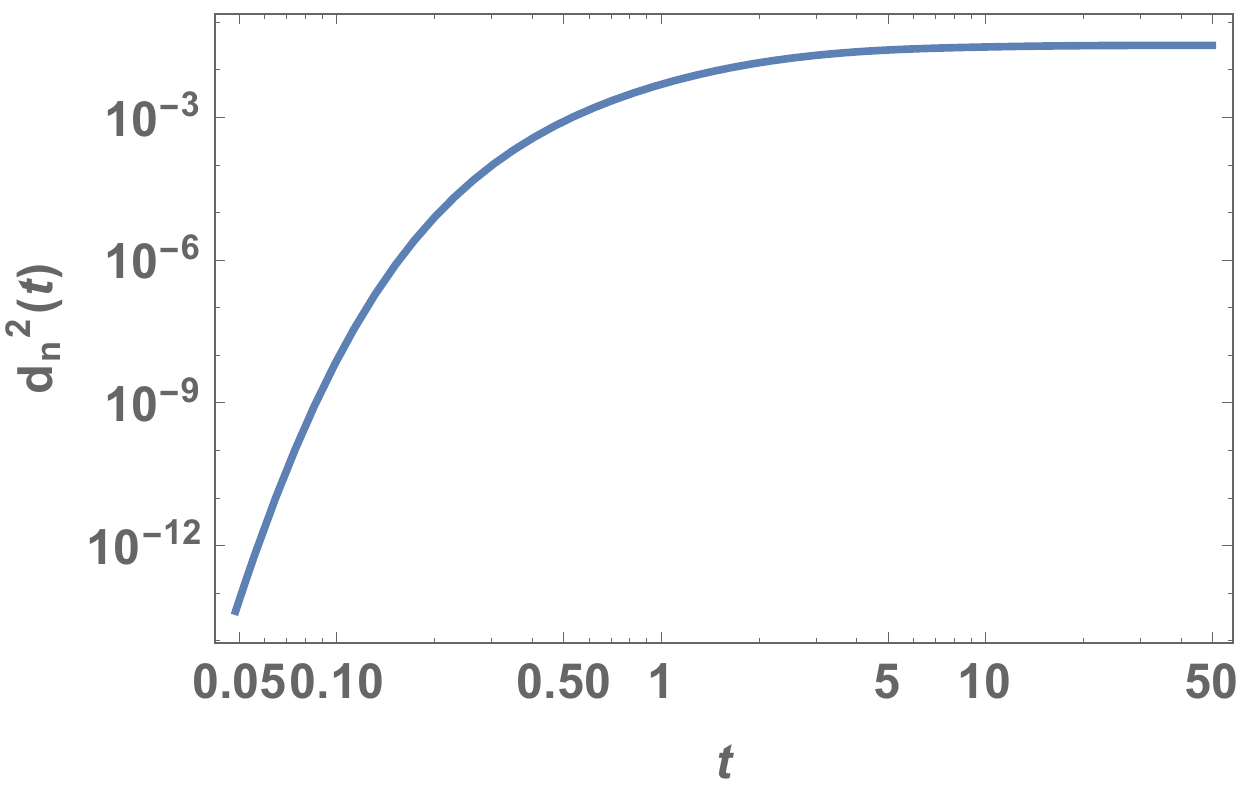}
\caption{Active contribution to the mean squared displacement of a monomer in a chain of $N=256$ where $12$ pairs of monomers feel an active force. The left (right) figure shows the behavior for a monomer that is (not) subject to an active force. The dashed red line represents a best fit through the data for $0.2<t<1$ and has a slope $0.22$. The data in the right figure show no evidence for a power law approach to the NESS.}
\label{fig7}
\end{figure}

In the case of heterogeneous active forces, the behavior of monomers is different whether they experience an active force or not. We focus again on the active contribution to $d_n^2(t)$.
For a monomer that doesn't feel active forces, the behavior is not power law (see Fig. \ref{fig7}, right hand side). For a monomer that feels an active force, there is a power law regime but the power is lower than $1/2$ and increases with the fraction of monomers subject to the active force. For example, at a density of active sites of $0.1$, the exponent is $0.22$ (see Fig. \ref{fig7}, left hand side) whereas at a density of $0.5$ it equals $0.39$. 
To this active contribution we have to add the thermal one. However, since the active forces are in general much stronger than the thermal ones, we find that the exponent is only slightly modified. 
We mention here that in a study of telomeres in mammalian cells,  a subdiffusion with an exponent of $0.32$ was found \cite{Bronstein09} while in a recent study a somewhat higher value of $0.41$ was found \cite{Stadler17}.

In Fig. \ref{fig7} we also notice that there is a large heterogeneity in the displacements of various monomers. At the time of around one second they differ by two orders of magnitude. Such behavior has also been observed experimentally \cite{Javer13,Stadler17} . 
In any case, we do not see any evidence for the superdiffusive behavior present for the case of monopolar active forces \cite{Vandebroek15,Sakaue16}. 

The deviation of the exponent from $1/2$ observed experimentally has been attributed to the viscoelastic nature of the cellular environment \cite{Weber10,Polovnikov18}. It is possible to extend our present calculations to the viscoelastic case but here we choose not to add this extra complication. Our results suggest that both viscoelasticity and heterogeneity can contribute to the measured exponent. 

\subsection{Velocity autocorrelation function}
The instantaneous velocity of a monomer cannot be easily determined in an experiment. Following earlier work \cite{Lampo16}, we introduce therefore the average velocity of a monomer over a time $\delta$ as $\vec{v}_n(t,\delta)=(\vec{R}_n(t+\delta) - \vec{R}_n(t))/\delta$. We are interested in the autocorrelation function of this velocity $C^{vv}_n(t,\delta,\tau)=\langle \vec{v}_n(t+\tau,\delta) \cdot \vec{v}_n(t,\delta) \rangle$. It is a quantity that can be measured experimentally by tracking the position of a chromosomal locus as a function of time and which therefore has been used as an important probe to determine the nature of the stochastic process underlying the observed motion \cite{Lampo16,Lucas14,DiPierro18}. 

The velocity autocorrelation function can, by definition, be expressed in terms of position autocorrelations, which within our model can be determined from that of the Rouse modes. The calculation is straightforward. We obtain

\begin{eqnarray}
C^{vv}_n(t,\delta,\tau)  &=&\frac{1}{\delta^2} \Big[ \frac{6 k_B T}{N \gamma} \left[\delta - \min(\tau,\delta)\right]\nonumber \\ +  \frac{6 k_B T}{N \gamma}&&\!\!\!\! \sum_{p=1}^{N-1} (C_n^p)^2\ \tau_p\ F_p(\delta,\tau) \label{velcor} \\
+ \frac{4 f^2 \lambda_0}{\gamma^2 N^2(\lambda_0+\lambda_f)} && \!\!\!\!\!\!\!\!\sum_{p=1}^{N-1} \sum_{q=1}^{N-1}  C_n^p C_n^q G_{p,q} F_{p,q}(t,\delta,\tau) \Big] \nonumber
\end{eqnarray}
where
\begin{eqnarray}
F_p(\delta,\tau) = 2 e^{-\tau/\tau_p} - e^{-(\tau+\delta)/\tau_p} - e^{-|\tau-\delta|/\tau_p}
\end{eqnarray}
and 
\begin{eqnarray}
F_{p,q}(t,\delta,\tau)&=& H_{p,q}(t+\tau+\delta,t+\delta) - H_{p,q}(t+\tau+\delta,t) \nonumber \\ &-& H_{p,q}(t+\tau,t+\delta) + H_{p,q}(t+\tau,t)
\end{eqnarray}
The first two terms are the thermal contribution which does not depend on $t$. Their behavior has been thoroughly analysed before. For $\delta \ll \tau_1$, where the monomer subdiffuses with an exponent $1/2$, the velocity autocorrelation function is well approximated by that of a point particle performing fractional Brownian motion (fBm) with exponent $1/2$ which equals \cite{Mandelbrot68}
\begin{eqnarray}
\frac{C^{vv}_{\textnormal{fBm}}(\delta,\tau)}{C^{vv}_{\textnormal{fBm}}(\delta,0)}=\frac{|\tau-\delta|^{1/2} + |\tau+\delta|^{1/2} - 2 |\tau|^{1/2}}{2 \delta^{1/2}}
\end{eqnarray}
In the opposite regime, $\delta \gg \tau_1$, the monomer follows the ordinary diffusion of the centre of mass for which the velocity autocorrelation is dominated  by the first term in (\ref{velcor}) which is proportional to $\delta - \tau$ for $\tau < \delta$ and zero for $\tau> \delta$. In both regimes there appears a negative peak in the autocorrelation at $\delta=\tau$. 

If we add active forces to a system in equilibrium, $C_n^{vv}$ will also depend on $t$, but as the NESS is reached this dependence should disappear again. Here we only present results for this latter case since this should also correspond to the situation for experiments that are performed {\it in vivo}. We also focus on the situation of homogeneous dipolar active forces.

\begin{figure}
\centering
\includegraphics[width=4cm]{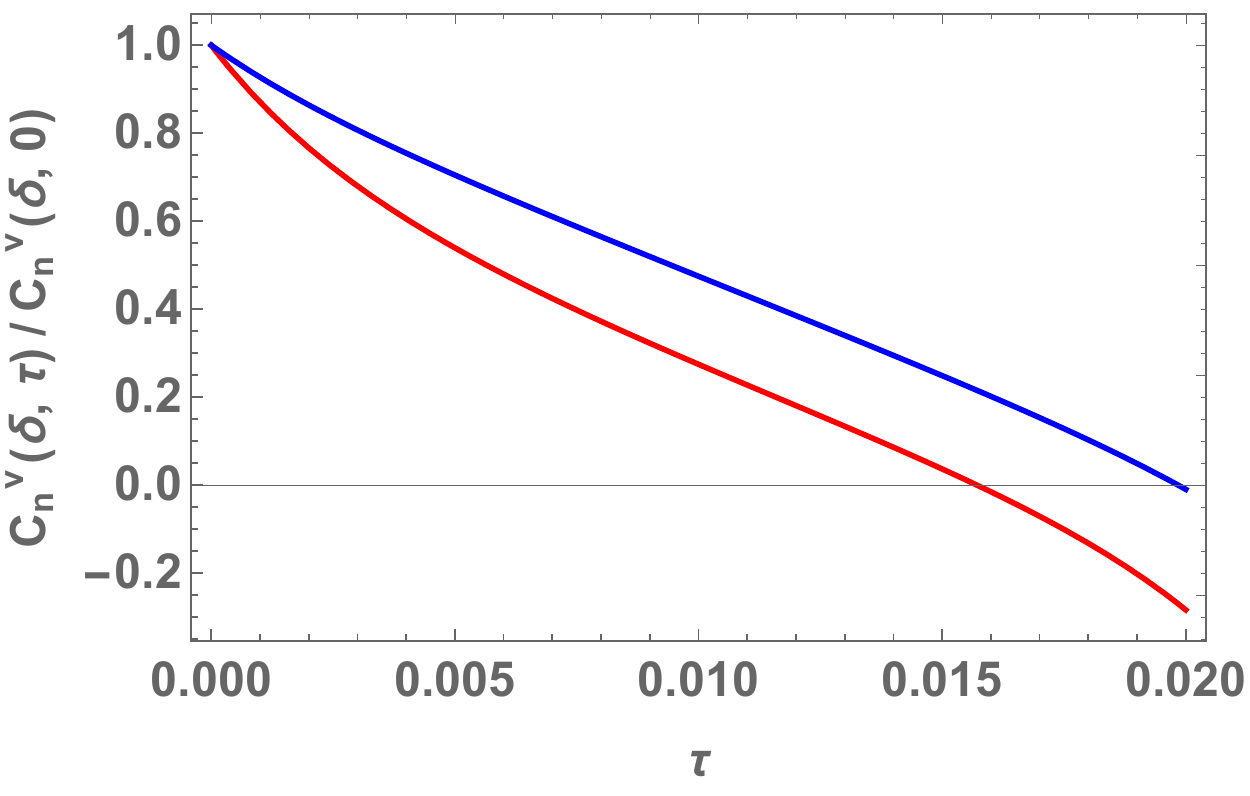}
\includegraphics[width=4cm]{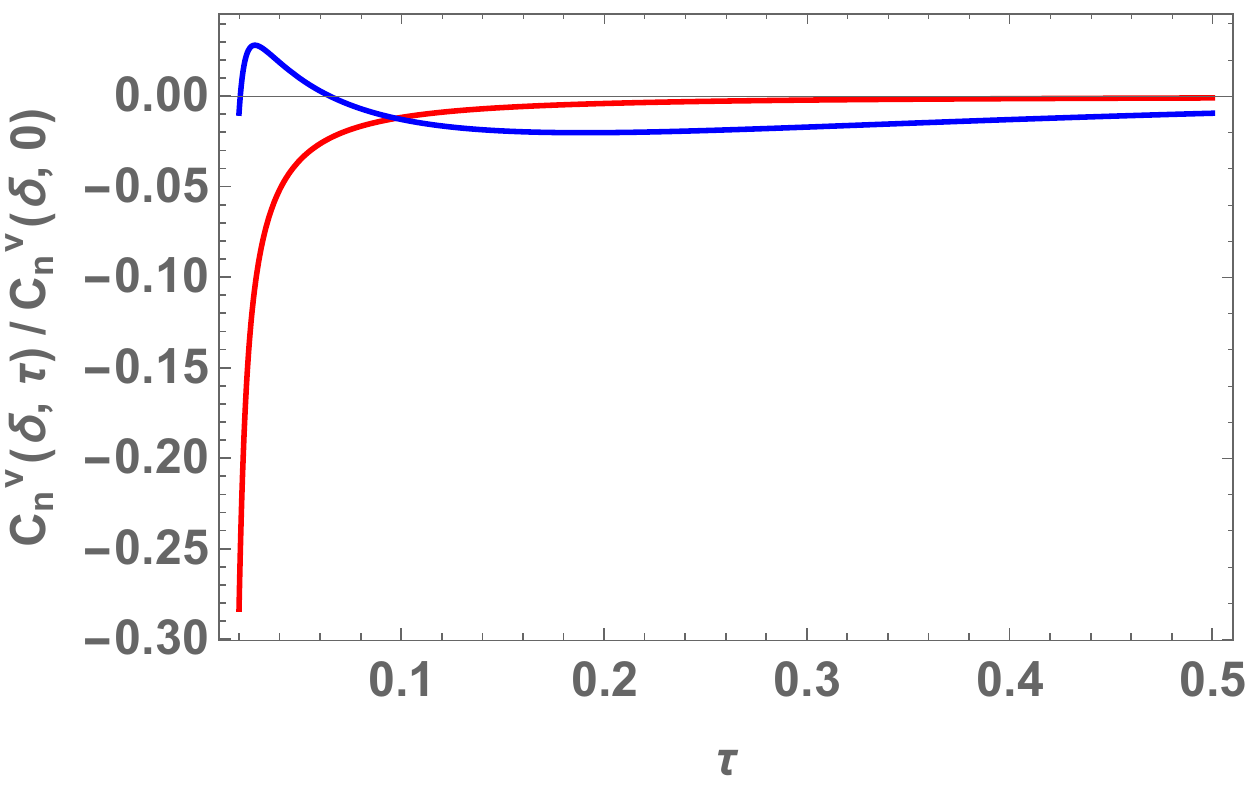}
\caption{Velocity autocorrelation as a function of $\tau$ for $\delta=0.02$ s and in the presence/absence of active forces (blue/red curve). Here $N=128$, $\tau_1 \approx 13$ s and $\tau_f=0.2$s. The diagram on the left(right) are the results for $\tau < \delta$ ($\tau > \delta$).}
\label{fig9}
\end{figure}

In Fig. \ref{fig9} we have plotted (blue line) $C^v_n(\delta,\tau)\equiv C^{vv}_n(\infty,\delta,\tau)$ for the middle monomer in a chain with $N=128$ as a function of $\tau$ for  $\delta < \tau_f \ll \tau_1$ , this is the time regime for which we found that the monomer performs a subdiffusion with exponent $1/2$ but with an enhanced diffusion constant. 
For clarity, we have made different plots for $\tau<\delta$ (left) and $\tau>\delta$ (right). Clearly, the velocity autocorrelation is completely different from that in the thermal case (red curves). Firstly, the negative peak has almost disappeared. Moreover, there appears a positive peak and a broader minimum which is located near $\tau\approx \tau_f$. This is strong evidence that while the particle performs a subdiffusion, the underlying stochastic process is not fractional Brownian motion. 

\begin{figure}
\centering
\includegraphics[width=8cm]{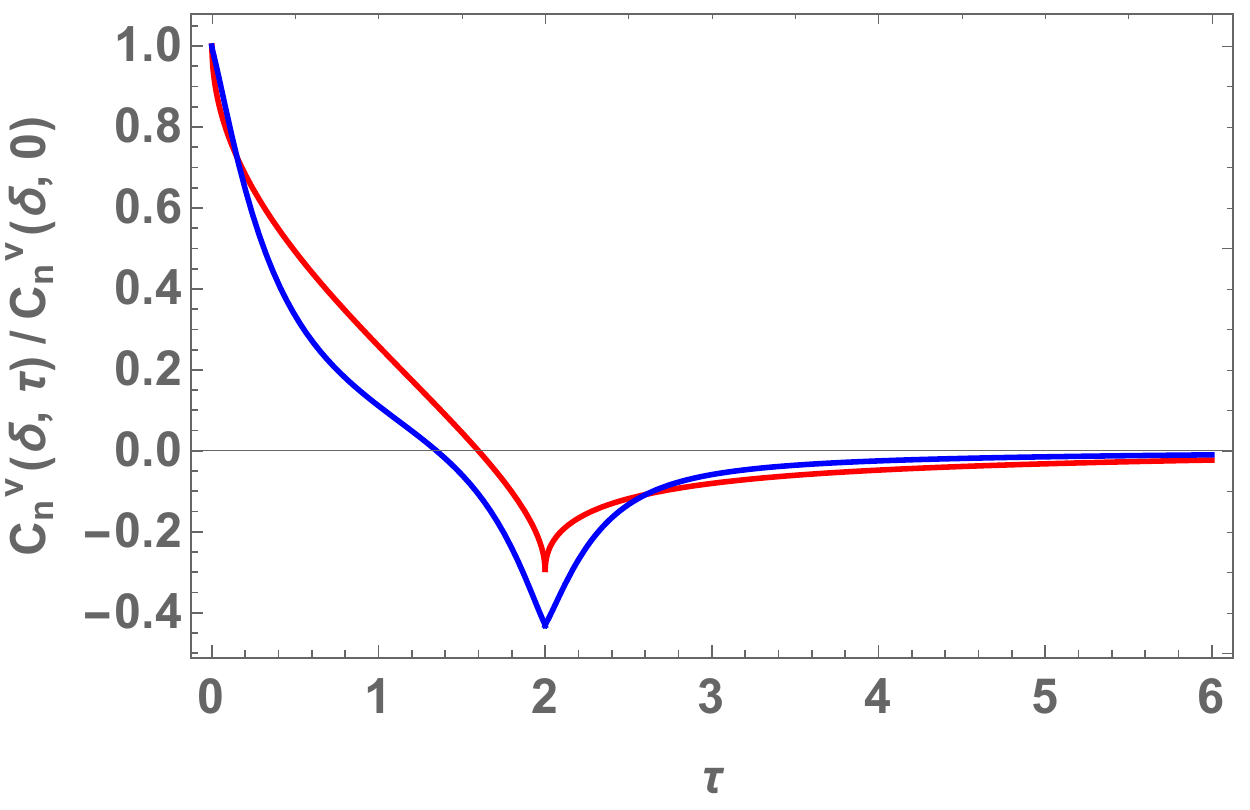}
\caption{Velocity autocorrelation as a function of $\tau$ for $\delta=2$ s and in the presence/absence of active forces (blue/red curve). Here $N=128$, $\tau_1 \approx 13$ s and $\tau_f=0.2$s.}
\label{fig10}
\end{figure}

For a value of $\delta$ such that $\tau_f < \delta \ll \tau_1$, we have plotted the velocity autocorrelation function in Fig. \ref{fig10} for the active case (blue) and the thermal one (red). There are differences between the two curves but qualitatively they are similar. The initial decrease is faster in the active case and there is again a pronounced negative peak. 

\begin{figure}
\centering
\includegraphics[width=8cm]{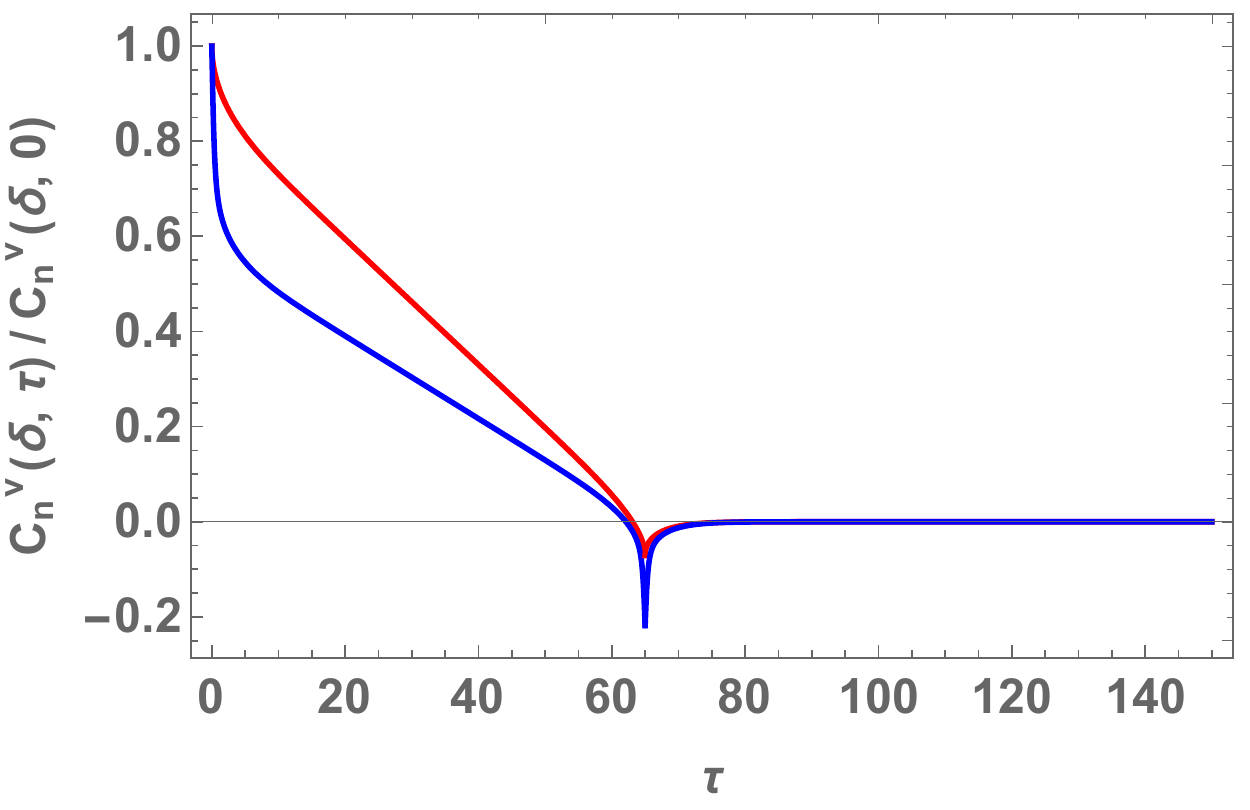}
\caption{Velocity autocorrelation as a function of $\tau$ for $\delta=65$ s and in the presence/absence of active forces (blue/red curve). Here $N=128$, $\tau_1 \approx 13$ s and $\tau_f=0.2$s.}
\label{fig11}
\end{figure}

Finally, for $\delta \gg \tau_1$ (Fig. \ref{fig11}), the difference between the active and thermal behavior is concentrated at low $\tau$ after which also the active contribution decreases linearly. For $\tau>\delta$ both curves almost coincide. 

In conclusion, we find that the influence of active forces on the velocity autocorrelation function shows up in the regime where $\tau \leq \tau_f$. If $\delta$ is also in this time window this leads to a qualitatively different behavior. For $\delta$ larger, the difference between the two cases remains but is more quantitative. 

We are not aware of any experimental evidence for the behavior that we found here, but this could be because the measurements where not made in the correct time window.

Clearly, also in the case of polymers subject to active forces, the velocity autocorrelation function turns out to be an important tool for the analysis of the stochastic processes involved.

\section{Monopolar superdiffusion versus dipolar subdiffusion}
In this section we clarify why active dipolar forces lead to subdiffusive motion, and this in contrast to the superdiffusive behavior that was found for monopolar forces. The argument is an extension of that in \cite{Sakaue16} and is based on the diffusive spreading of stress in a Rouse chain \cite{Sakaue12}. 

In the continuum limit, the equation of motion of the $n$-th monomer becomes a non-homogeneous diffusion equation
\begin{eqnarray}
\frac{\partial \vec{R}(n,t)}{\partial t} = \frac{k}{\gamma} \frac{\partial^2 \vec{R} (n,t)}{\partial n^2} + \vec{F_a}(n,t)
\label{cont}
\end{eqnarray}
where $\vec{F_a}(n,t)$ is an arbitrary time dependent force acting on the $n$-th monomer. The solution of this equation can be written in terms of the Green function $G(n,t)$ which is the solution to (\ref{cont}) for the case that $F_a(n,t)=\delta(t) \delta_{n,0}$
\begin{eqnarray}
\frac{\partial G(n,t)}{\partial t} = \frac{k}{\gamma} \frac{\partial^2 G(n,t)}{\partial n^2} + \delta(t) \delta_{n,0}
\end{eqnarray}
For free boundary conditions at both chain ends, one has
\begin{eqnarray}
G(n,t) = \left( \frac{\gamma}{4\pi k t}\right)^{1/2} \exp\left(-\frac{\gamma}{4kt} n^2 \right)
\label{Green}
\end{eqnarray}

In terms of this Green function, the solution to (\ref{cont}) can be written as
\begin{eqnarray}
\!\!\!\!\!\!\!\!\langle \Delta \vec{R}(n,t) \rangle = \frac{1}{\gamma} \sum_{m=0}^{N-1} \int_0^t dt' G(n-m,t-t') \vec{F}_a (m,t')
\end{eqnarray}
where we have made a thermal average over the initial positions. 

Let us look at the response to a monopolar force acting only on monomer $n$. It is given by
\begin{eqnarray}
\langle \Delta \vec{R}(n,t) \rangle = \frac{1}{\gamma} \int_0^t dt' G(0,t-t') \vec{F_a}(n,t')
\label{vgl5}
\end{eqnarray}
The Green function is $G(0,t)=(\gamma/4\pi k t)^{1/2}$. For a constant force $\vec{F}_a(n)$, the displacement then grows as $t^{1/2}$. Therefore, Eq. (\ref{vgl5}) can be recast as
\begin{eqnarray}
\Gamma(t) \frac{\partial \langle \vec{R}(n,t)\rangle}{\partial t} = \vec{F}_a(n)
\end{eqnarray}
with $\Gamma(t) \sim t^{1/2}$, i.e., the effective friction for the motion of a tagged monomer increases as the tension propagates along the chain. 

In the case of a dipolar active force acting on the monomers $n$ and $n+1$, we similarly have
\begin{eqnarray*}
\langle \Delta \vec{R}(n,t) \rangle = \frac{1}{\gamma} \int_0^t dt' [G(0,t-t')-G(-1,t-t')] \vec{F}_a(n,t') \nonumber \\
\end{eqnarray*}
We now apply the expansion of the Green function with respect to the chain coordinate $n$
\begin{eqnarray*}
G(n+\delta n,t) = G(n,t) + G'(n,t) \delta n + \frac{1}{2} G''(n,t) \delta n^2 + \cdots
\end{eqnarray*}
Since $G'(n,t)|_{n=0}=0$ and $G''(n,t)|_{n=0}=-(2\pi)^{-1/2} (\gamma/(2kt))^{3/2}$, we obtain
\begin{eqnarray}
\langle \Delta \vec{R}(n,t) \rangle = \frac{1}{\gamma} \int_0^t dt' G_d(t-t') \vec{F}_a(n,t')
\end{eqnarray}
with
\begin{eqnarray}
G_d(t) = \left( \frac{1}{8 \pi} \right)^{1/2} \left(\frac{\gamma}{2kt}\right)^{3/2} + {\cal O} \left(\left(\frac{\gamma}{kt}\right)^{5/2}\right)
\end{eqnarray}
Going through a similar argument as for the monopolar force, we find that the effective friction now grows as $t^{3/2}$. Note that tension still propagates as $t^{1/2}$, that is a definite property of the Rouse chain, but the dipolar configuration of the force modifies the scaling for the effective friction $\Gamma(t) \sim t^{1/2} \to t^{3/2}$. 

We now turn to the mean squared displacement. In \cite{Sakaue16} is has been shown that $d_n^2(t)$ (for a homogeneous chain) can be expressed in terms of $\Gamma(t)$ and the mean squared displacement $\langle \Delta \vec{R}_\star(t)^2\rangle$  of an unconnected monomer (i.e. a non-polymeric free particle) subject to the same (thermal or active) random forces,
\begin{eqnarray}
d_n^2(t) = \Gamma^{-1} (t) \langle \Delta \vec{R}_\star(t)^2\rangle
\label{vgl1}
\end{eqnarray}
In the time regime $t < \tau_f$, the latter displays a ballistic behavior, thus $\langle \Delta \vec{R}_\star(t)^2\rangle \sim t^2$ \cite{Vandebroek17b}. Plugging this in (\ref{vgl1}) together with the results obtained above for $\Gamma(t)$ shows that for the monopolar active forces, we have a superdiffusive scaling $d_n^2(t) \sim t^{3/2}$ consistent with the result of \cite{Vandebroek15}. On the other hand, for the dipolar active forces we obtain $d_n^2(t) \sim t^{1/2}$, i.e. the subdiffusive behavior derived in the previous section. 

Although the final $t^{1/2}$ scaling is the same as the conventional result on the Rouse chain in a purely thermal environment, we see that the underlying physics is very different.

\section{Monomer-monomer correlation function and the appearance of coherent motion}
Recent advances in experimental techniques have allowed to measure the correlation of motion in chromatin throughout the whole nucleus \cite{Zidovska13,Shaban18,Saintillan18}. It has been observed that active processes lead to coherent motion over regions of several $\mu$m. To see whether such behavior also shows up in our model we have calculated the correlation $D_{n,m}(t)$ in the motion of two monomers, $n$ and $m$. We find
\begin{eqnarray}
D_{n,m}(t)&=& \langle \Delta \vec{R}_n(t) \cdot \Delta \vec{R}_m(t)\rangle=\frac{6 k_B T}{\gamma N} t \nonumber \\ &+& 24k_B T \sum_{p=1}^{N-1} C_n^p C_m^p  \left(1- \exp(-t/\tau_p)\right)/k_p \label{corr2} \\
&+& \frac{4 \lambda_0 f^2}{\gamma^2 N^2 (\lambda_0+\lambda_f)} \sum_{p=1}^{N-1}\sum_{q=1}^{N-1} C_n^p C_m^q G_{p,q} H_{p,q}(t,t) \nonumber
\end{eqnarray}

\begin{figure}
\centering
\includegraphics[width=8cm]{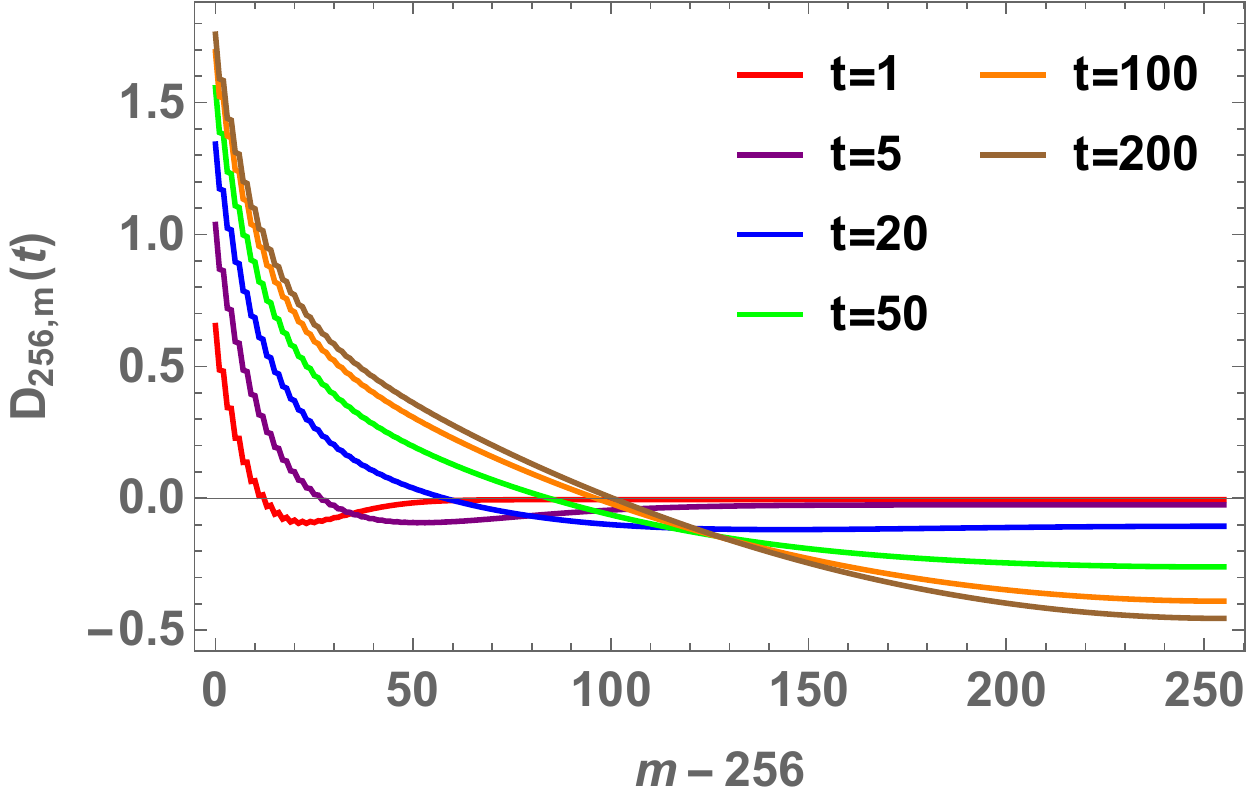}
\caption{Correlation $D_{n,m}(t)$ between the motion of monomer $m$ and the middle monomer in a chain with $N=512$ for various times (homogeneous dipolar forces).}
\label{fig12}
\end{figure}

Firstly, we investigate the case where all monomers can bind enzymes, i.e. to the case of homogeneous active forces. We disregard the motion of the centre of mass. 
In Fig. \ref{fig12} we show, for various times, the correlation between monomer $m$ and the middle monomer in a chain with $512$ monomers. We see that as a function of time the motion becomes more coherent in the sense that for monomers close to the middle one the correlation increases, while for those far away the correlation becomes more negative. This indicates that the first group tends to move more parallel to the middle one, whereas the second group moves more antiparallel. 

\begin{figure}
\centering
\includegraphics[width=8cm]{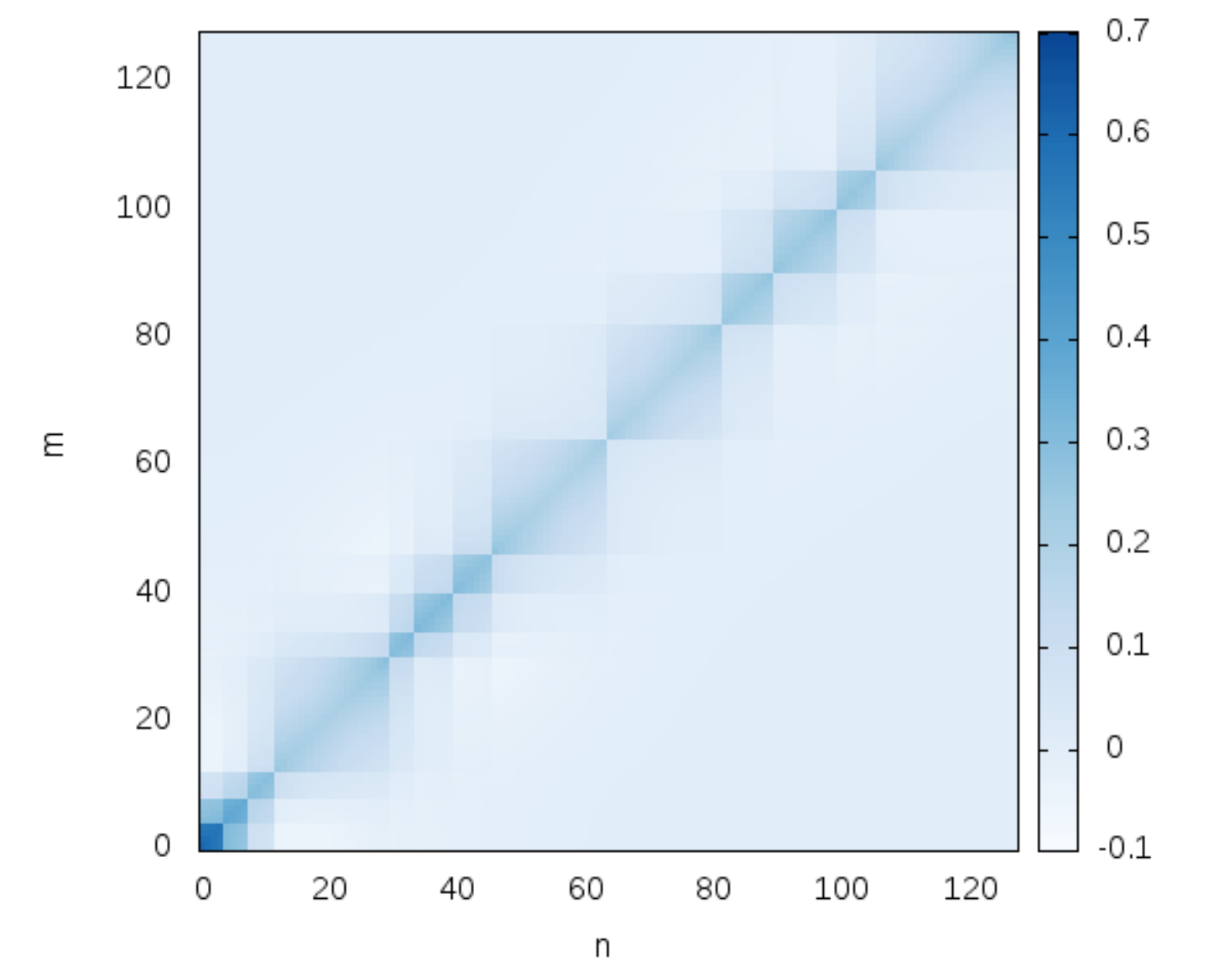}
\includegraphics[width=8cm]{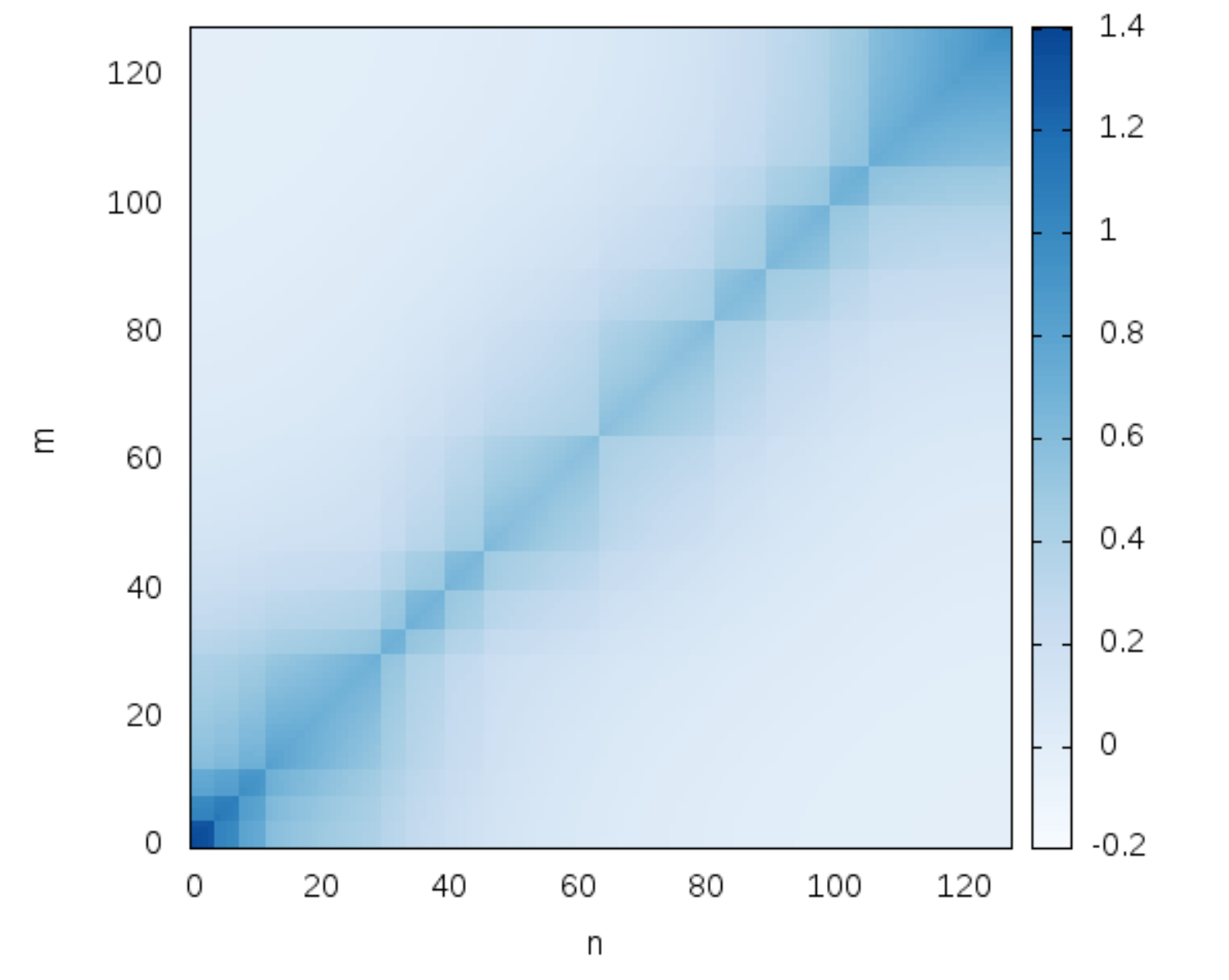}
\caption{Correlation matrix $D_{n,m}(t)$ (in $\mu$m$^2$) in a chain with $N=128$ at $t=1$ (top) and $t=10$ (bottom) seconds. Active dipole forces are applied at 12 monomer pairs. }
\label{fig14}
\end{figure}

A similar increase in coherence is observed for the case of heterogeneous active interactions. In Fig. \ref{fig14}  we show the whole matrix $D_{n,m}(t)$ at $t=1$ and $t=10$ seconds in a chain of $N=128$ where active dipoles are present at 12 randomly chosen pairs of monomers. One observes the appearance of regions of coherent motion. The boundaries of these regions are the monomers where the active forces act. Within each of these regions the correlation increases as a function of time while the correlations between monomers that are far apart along the chain become more negative. The correlations become time independent when the NESS is reached.

Coherent motion has also been found in polymer models without active forces \cite{DiPierro16,DiPierro18}. In these models one includes interactions between monomers that can be far apart along the chain. The precise form of these can, for example, be obtained from experimental data from Hi-C measurements \cite{Lieberman09}. The extra interactions lead to the formation of loops. One then finds that the motion becomes coherent inside the loops. 
The advantage of these models is that, by construction, they reproduce the structure of chromatin as determined from Hi-C experiments. However, we believe that active forces must play an important role in the  dynamical behavior of chromatin since 
in experiments one observes that the dynamical coherence disappears when ATP is diluted \cite{Zidovska13}. It would therefore  be of interest to determine the motion in a polymer model which includes both loops and active forces. 

\section{Mean squared end-to-end distance}
The squared end-to-end distance, $R^2(N,t)$, of the polymer is given by $(\vec{R}_{N-1}(t)-\vec{R}_0(t))^2$. Using (\ref{monpos}) and assuming that $N \gg 1$, we get the standard relation for the average squared end-to-end distance
\begin{eqnarray}
\langle R^2 (N,t) \rangle = 16 \sum_{p,q\atop \textnormal{odd}} \langle \vec{X}_p(t) \vec{X}_q(t) \rangle
\end{eqnarray}
When inserting (\ref{eq:XpXq}) one notices that there is a contribution from thermal and from active forces. The former equals $3 k_B T N/k$ which can easily be understood on the basis of the equipartition theorem \cite{Doi86}. 

We therefore focus our discussion on the effect of the active forces. 
Using (\ref{eq:XpXq}) one finds that the active contribution  to the average squared end-to-end distance equals 
\begin{eqnarray}
\frac{16 \lambda_0}{\lambda_0 + \lambda_f} \frac{f^2}{N^2 \gamma^2} \sum_{p,q\atop \textnormal{odd}} G_{p,q} H_{p,q}(t,t)
\label{sizeactive}
\end{eqnarray}
At very early times ($t \ll \tau_p, t \ll \lambda_f^{-1}$), the exponentials in $H_{p,q}(t,t)$ are to good approximation equal to $1$ so that the active contribution (\ref{sizeactive}) initially grows as $t^2$. Secondly, the integrals (\ref{integrals}) are increasing functions of time which asymptotically  reach the value
\begin{eqnarray}
H_{p,q}^\star &=&\lim_{t \to \infty} H_{p,q}(t,t)\nonumber \\ &=& \dfrac{\tau_p \tau_q (\tau_p + \tau_q + 2\lambda_\textnormal{f}\tau_p \tau_q)}{(1+\lambda_\textnormal{f}\tau_p)(1+\lambda_\textnormal{f}\tau_q)(\tau_p+\tau_q)}.
\label{asymp}
\end{eqnarray}
In terms of these quantities, the mean squared end-to-end distance of the polymer in the NESS equals
\begin{eqnarray*}
\langle R^2 (N,t \to \infty) \rangle &=& \frac{3 k_B T}{k} N\\ &+& 
\frac{16 \lambda_0}{\lambda_0 + \lambda_f} \frac{f^2}{N^2 \gamma^2} \sum_{p,q\atop \textnormal{odd}} G_{p,q} H_{p,q}^\star
\end{eqnarray*}

For the case in which the enzymes can bind to all monomers ($G_{p,q}$ is then given by (\ref{eq:Gpq})), it can be shown (see supplemental material) that for large $N$ the active contribution to $\langle R^2 (N,t) \rangle$ reaches a constant. Hence, the polymer swells with an amount that is independent of $N$. This has to be contrasted to the case of a polymer subject to  monopolar active forces where this swelling is proportional to $N$ \cite{Vandebroek15,Sakaue16}. Our results are qualitatively consistent with experiments on interphase chromatin which was found to condense after depletion of ATP \cite{Zidovska13}. The fact that the swelling does not depend on the number of monomers may have biological relevance because it could help to limit the size of a long chain. 

\begin{figure}
\centering
\includegraphics[width=4cm]{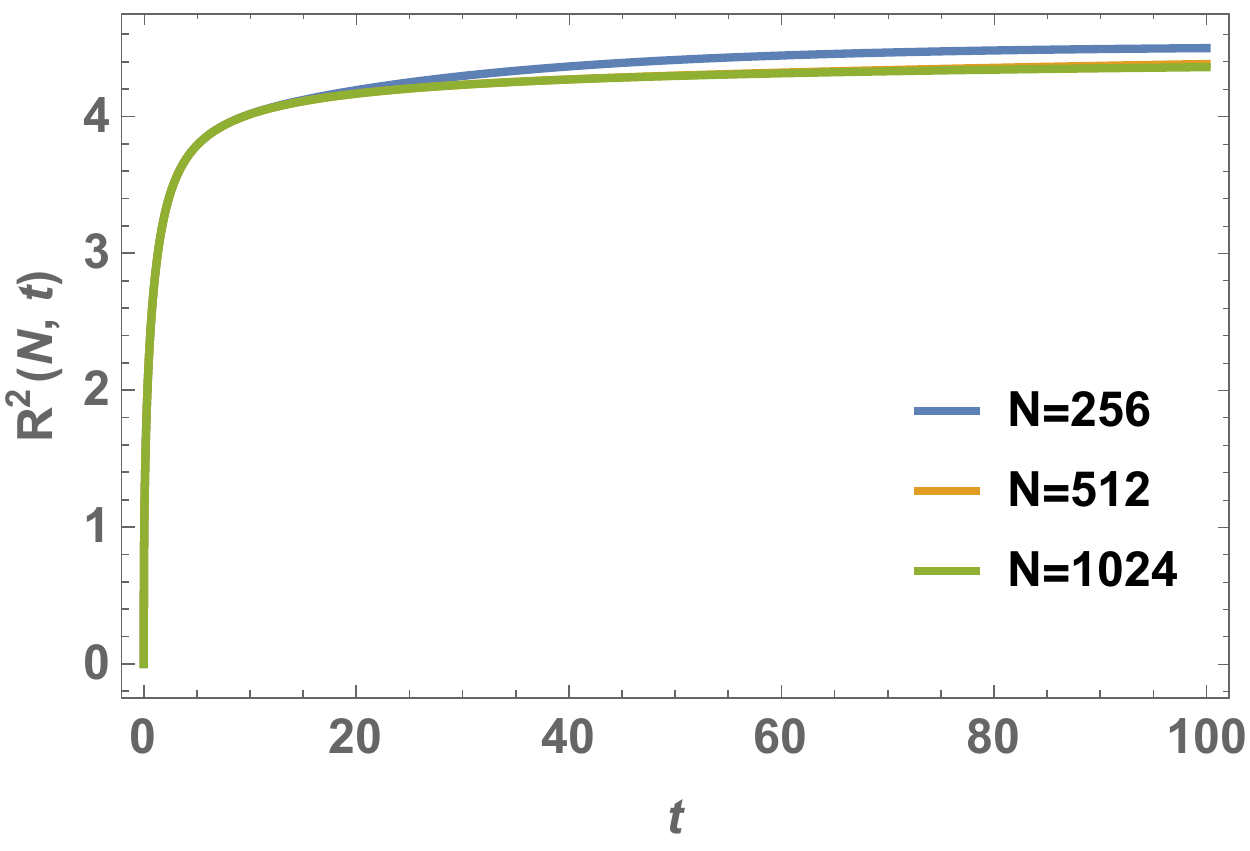}
\includegraphics[width=4cm]{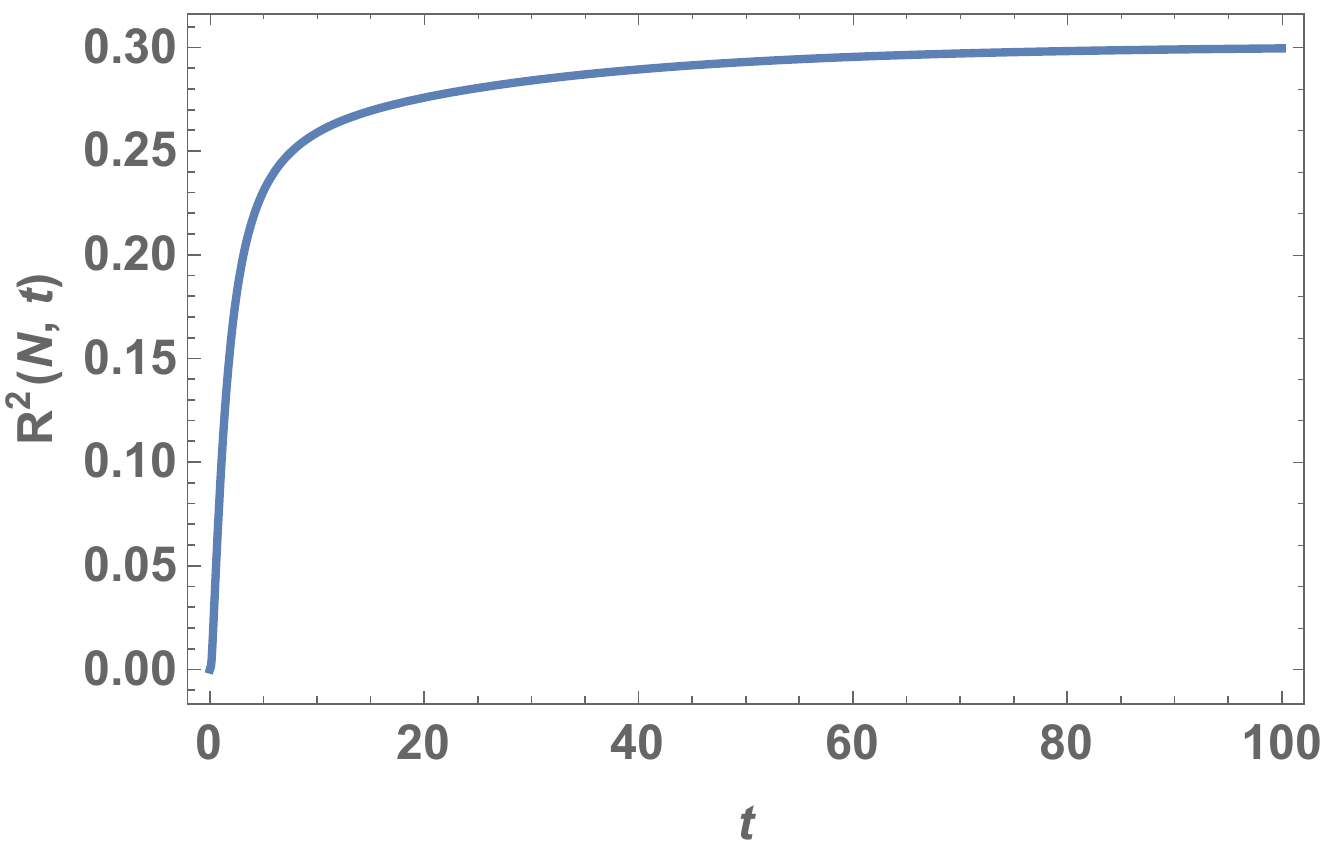}
\caption{Time dependence of the average squared end-to-end distance versus time. Only the active contribution is shown. On the left hand side we show the results for a chain where enzymes can bind on all sites (for $N=256, 512$ and $1024$). On the right hand side we show a case where there are only $12$ binding sites chosen at random positions in a chain of $N=256$.}
\label{fig3}
\end{figure}

In Fig. \ref{fig3} we show the time dependence of the mean squared end-to-end distance for chains with $N=256, 512$ and $N=1024$ (only the active contribution). On the left side of the figure, we see that for the case where enzymes can bind on all the monomers, the increase in size does not depend on $N$ for $N$ large. We see that after a time of the order of 10 seconds the size of the polymer remains constant indicating that the NESS is reached. In total, the mean squared end-to-end distance increases with a few micrometer. On the right hand side of Fig. \ref{fig3}, we show an example where dipolar forces are only acting on 24 of the monomers in a polymer with $N=256$. The data show that the behavior of $\langle R^2(N,t)\rangle$ is qualitatively the same as in the previous case and indicate that the increase in size in the NESS is proportional to the fraction of binding sites.

\section{Conclusions}
In this paper we have studied a solvable model of a flexible polymer that interacts with active enzymes, i.e. proteins that convert the free energy released by ATP hydrolysis into force and motion. Our model was introduced with the dynamics of chromosomes as an inspiration. However, it could also be relevant for other situations, for example the interaction between a chaperone and a protein. 

We focused on the motion of individual monomers and have found them to be subdiffusive. This result can also be understood on the basis of stress propagation through a Rouse chain. It is an essential feature of dipolar forces. From a numerical study, we find that the exponent of the subdiffusion depends on the density of active enzymes. Heterogeneity is therefore, next to viscoelasticity of the cellular environment, a possible cause for the subdiffusive exponent to differ from that in the equilibrium Rouse chain. 

We have also determined the velocity autocorrelation function and have found that it shows clear signatures of the active processes. Its form is different from that of fractional Brownian motion. This is no surprise. It is possible to derive the equation of motion of a tagged monomer in an equilibrium Rouse chain by eliminating the position of all the other monomers. The resulting equation is a generalised Langevin equation (GLE) with a memory kernel that is power law up to the Rouse time \cite{Panja10,Sakaue13,Maes13}. The noise term is related to the kernel by the fluctuation-dissipation theorem (FDT). From this GLE one deduces that the monomer subdiffuses with an exponent $1/2$ up to the Rouse time and crosses over to ordinary diffusion afterwards. While the generalised Langevin equation and fractional Brownian motion describe different stochastic processes, they are closely related \cite{Jeon10}. In a similar spirit,  the equation of motion for a monomer in a Rouse chain subject to monopolar active forces was recently derived \cite{Vandebroek17}. In the resulting equation the memory kernel is not modified with respect to equilibrium but an extra noise term appears that breaks the FDT. The resulting equation of motion is therefore no longer simply related to fBm. In principle it should also be possible to derive the equation of a tagged monomer for the case of dipolar forces. In practice that is a difficult calculation, especially if the forces are not homogeneous and the Rouse modes are not real eigenmodes.
It would be of interest to see if the behavior of the velocity autocorrelation derived here is also present in more realistic polymer models and in experiment.

We have given clear evidence for the appearance of regions of correlated motion. This phenomenon has been observed in recent experiments. In a very recent work \cite{Saintillan18}, it was shown numerically to occur in a polymer model where  the dipolar forces do not act directly on the chain but influence the hydrodynamics of the environment, which then in turn modifies the friction experienced by each monomer. In our model, the dipolar forces act directly on the polymer chain by bringing it out of equilibrium. Their effect then spreads through the polymer by tension relaxation. 

While our model is an oversimplification for the real chromosome, we have seen that its behavior shows many intriguing similarities with those found in experiments. Our work suggests that heterogeneous subdiffusion and correlated motion (and possibly also the behavior seen in the velocity autocorrelation) are generic properties of polymers subject to dipolar active  forces. It is certainly of interest to see whether they also appear in more realistic polymer models which also take into account the formation of loops through long range interactions along the chain \cite{DiPierro16,Nuebler18}. From our work and that of others, it is however already clear that the observed dynamical behavior is a universal byproduct of the active processes that go on inside a cell. Whether the resulting complex dynamics also plays a biological role in, for example, gene expression remains an open question.
  
\begin{acknowledgements}
T.S thanks T. Yamamoto (Nagoya University) for discussion on polymer
dynamics, and financial support from JSPS KAKENHI (No. JP18H05529)
from MEXT, Japan, and JST, PRESTO (JPMJPR16N5).
\end{acknowledgements}

\end{document}